\documentclass[12pt]{article}
\usepackage{graphicx}
\newcommand\pubdate{\today}
\newcommand\pubnumber{EFI 11-17}
\def\Title#1{\begin{center} {\Large #1 } \end{center}}
\def\Author#1{\begin{center}{ \sc #1} \end{center}}
\def\Address#1{\begin{center}{ \it #1} \end{center}}

\newcommand\pubblock{\rightline{\begin{tabular}{l} \pubnumber\\
         \pubdate  \end{tabular}}}
\newenvironment{Abstract}{\begin{center}{\bf Abstract}\end{center} \bigskip \begin{quotation}  }{\end{quotation}}
\newenvironment{Presented}{\begin{quotation} \begin{center} 
             PRESENTED AT\end{center}\bigskip 
      \begin{center}\begin{large}}{\end{large}\end{center} \end{quotation}}
\def\Acknowledgements{\bigskip  \bigskip \begin{center} \begin{large}
             \bf ACKNOWLEDGEMENTS \end{large}\end{center}}

\def\beq{\begin{equation}}
\def\eeq#1{\label{#1}\end{equation}}
\def\eeqn{\end{equation}}

\def\beqa{\begin{eqnarray}}
\def\eeqa#1{\label{#1}\end{eqnarray}}
\def\eeqan{\end{eqnarray}}

\let\bar=\overbar

\def\Dslash{\not{\hbox{\kern-4pt $D$}}}
\def\dslash{\not{\hbox{\kern-2pt $\del$}}}

\def\msb{{\bar{\ssstyle M \kern -1pt S}}}

\def \bea{\begin{eqnarray}}
\def \beq{\begin{equation}}
\def \brf{{\cal B}}
\def \eea{\end{eqnarray}}
\def \eeq{\end{equation}}
\begin{document}
\begin{titlepage}
\pubblock

\vfill


\Title{Quarkonium -- Theory}
\vfill
\Author{Jonathan L. Rosner}  
\Address{Enrico Fermi Institute, University of Chicago, Chicago, IL 60637, USA}
\vfill


\begin{Abstract}

Some recent issues in the theory of heavy quarkonium are discussed.  Many
of these deal with the need to extend the description of charmonium and
bottomonium states beyond the simple $Q \bar Q$ picture.  Some recent
progress on radiative transitions in bottomonium is also described.
\end{Abstract}

\vfill

\begin{Presented}
The Ninth International Conference on\\
Flavor Physics and CP Violation\\
(FPCP 2011)\\
Maale Hachamisha, Israel,  May 23--27, 2011
\end{Presented}
\vfill

\end{titlepage}
\def\thefootnote{\fnsymbol{footnote}}
\setcounter{footnote}{0}
%

\section{Introduction}

The discovery of the family of charmonium and bottomonium resonances in
the mid-1970s was greeted initially with simple descriptions in terms of
non-relativistic $c \bar c$ and $b \bar b$ bound states, illustrating basic
principles familiar from the earliest days of quantum mechanics
\cite{Quigg:1979vr,Grosse:1979xi}.  It has now come time to go beyond the
simple $Q \bar Q$ picture of heavy quarkonium.  Experiments have uncovered
new degrees of freedom (many associated with flavored mesons) and are now
at a level of accuracy sufficient to distinguish among many different
schemes of relativistic corrections.  We describe some of these recent
developments in this brief review.

Scalar mesons below 1 GeV (Sec.\ \ref{sec:scalars}) illustrate the importance
of coupled channels and mesonic degrees of freedom.  Another case in point,
known for 50 years, is the $\Lambda(1405)$ (Sec.\ \ref{sec:lambda}).  The
opening of new thresholds can lead to dips and cusps in mass spectra (Sec.\
\ref{sec:dips}); the interplay of closed and open channels is familiar from
Feshbach resonances in nuclear physics.

Recent discoveries of $Q \bar Q q \bar q$ exotic states, where $Q,q$ denote
heavy and light quarks, respectively, pose the question of whether these are
genuine tetraquark states or more closely associated with states of a flavored
meson $Q \bar q$ and antimeson $\bar Q q$.  Exotic baryonium states were
predicted 43 years ago (Sec.\ \ref{sec:bary}) but the jury is still out on
their existence.  One possibility for observing them is in $B$ meson
decays.  It appears that $Q \bar Q q \bar q$ states,
probably in the form of $B \bar B^*$ and $B^* \bar B^*$ ``molecules,''
play a key role in the recent observation of the decays $\Upsilon(5S) \to
\pi^+ \pi^- h_b(1P,2P)$ (Sec.\ \ref{sec:hb}), through rescattering from
states of open flavor.

We describe recent progress on radiative bottomonium transitions in Sec.\
\ref{sec:radt}, and conclude in Sec.\ \ref{sec:concl}.

A useful compendium of experimental references may be found in
\cite{Brambilla:2010cs}.  I draw heavily on the wisdom and common sense in
two articles by D. Bugg \cite{Bugg:2008kd,Bugg:2011ub}.

\section{Scalar mesons below 1 GeV \label{sec:scalars}}

The following candidates for positive-parity spinless mesons (see Ref.\
\cite{Nakamura:2010zzi} for a partial listing) probably owe
their existence to the mesonic channels to which they couple:

\begin{itemize}

\item $I=0$: The $\sigma(\sim 500)$, coupling to $\pi \pi$, is prominent in
many Dalitz plots.

\item $I=1/2$: The $\kappa(\sim 750)$, coupling to $K \pi$, also appears
frequently.

\item Another $I=0$ state, the $f_0(980)$, is closely correlated with the $K
\bar K$ threshold.

\item The $I=1$ state $a_0(980)$ couples to $\eta \pi$ and $K \bar K$.

\end{itemize}

All the properties of the above mesons are closely linked to coupled channels.
The $\sigma(500)$ is dynamically generated; it appears as a consequence of
current algebra, crossing, unitarity, and assumption of a $\rho$ in the
$I=J=1$ $\pi \pi$ channel \cite{Goble:1988cg}.  One expects similar dynamics to
generate a $\kappa$ in the $I=1/2$ $K \pi$ channel.

The $f_0(980)$ decays mainly to $\pi \pi$ but is produced largely from an $s
\bar s$ initial state, e.g., in $B_s \to J/\psi s \bar s$.  This behavior
was noticed quite early in $J/\psi$ decays:  The $f_0(980)$ appears in the
$\pi \pi$ spectrum in $J/\psi \to \phi \pi \pi$ but not $J/\psi \to \omega
\pi \pi$ \cite{Feldman:1977nj}.

A nonet structure (quark-diquark) has been proposed for the scalar mesons
below 1 GeV \cite{Jaffe:1977}.  However, it fails to describe quantitatively
the couplings of these states to meson-meson channels \cite{Bugg:2008kd}.

\section{Lessons from the $\Lambda(1405)$ \label{sec:lambda}}

The $\Lambda(1405)$ was originally identified 50 years ago as a low-energy
$I=0$ S-wave $\Sigma$-$\pi$ resonance \cite{Alston:1961zz}.  However, a key
feature is its strong coupling to the $I=0$ S-wave $\bar K N$ channel, whose
threshold lies $\sim 27$ MeV higher.  The interaction between closed and open
channels was studied extensively by Dalitz and Tuan in the late 1950s and early
1960s \cite{Dalitz:1959dn,Dalitz:1959dq,Dalitz:1960du} and represents a
realization of a {\it Feshbach resonance}, a phenomenon familiar from earlier
instances of nuclear physics \cite{Feshbach:1958nx}.  The opening of S-wave
channels such as the $I=0$ $\bar K N$ channel coupling to $\Lambda(1405)$
can lead to cusps and dips in scattering amplitudes.

The $\Lambda(1405)$ fits the SU(6) $\otimes$ O(3) quark model as a
$(70,~L=1~uds)$ state with $J^P = 1/2^-$.  Its large fine-structure splitting
from the state $\Lambda(1520)$ with $J^P = 3/2^-$ can be understood through
interactions with final kaon-nucleon and pion-hyperon final states
\cite{Isgur:1977ef,Isgur:1978xj}.  More recently, the $\Lambda(1405)$ has been
studied on the lattice \cite{Lage:2009zv} and recognized as a candidate for a 
$\bar K N$ molecule \cite{Hyodo:2011ur}.  It thus can be viewed both as a
conventional three-quark baryon and a meson-baryon composite.  An analogous
situation occurs for the $D_{s0}(2317)$, which can be viewed either as a
$^3P_0$ $c \bar s$ state (lower in mass than expected), or as a $KD$ state with
$~\sim 42$ MeV binding energy.

\section{Cusps and dips in mass spectra \label{sec:dips}}

Rapid variations in mass spectra are ubiquitous near S-wave thresholds
\cite{Bugg:2008kd,Wigner:1948,Rosner:2006vc}.  One sees cusps in the
$M(\pi^0 \pi^0)$ spectrum from $K_L \to 3 \pi^0$ decays at $\pi^+ \pi^-$
threshold \cite{Batley:2005ax}, permitting the measurement of the $\pi \pi$
S-wave scattering length difference $a_0-a_2$ \cite{Cabibbo:2005ez}.  Another
cusp is visible in $M(\pi^0 p)$ at $\pi^+ n$ threshold \cite{Schmidt:2001vg}.  

\begin{figure}
\hskip 0.8in \includegraphics[width=0.71\textwidth]{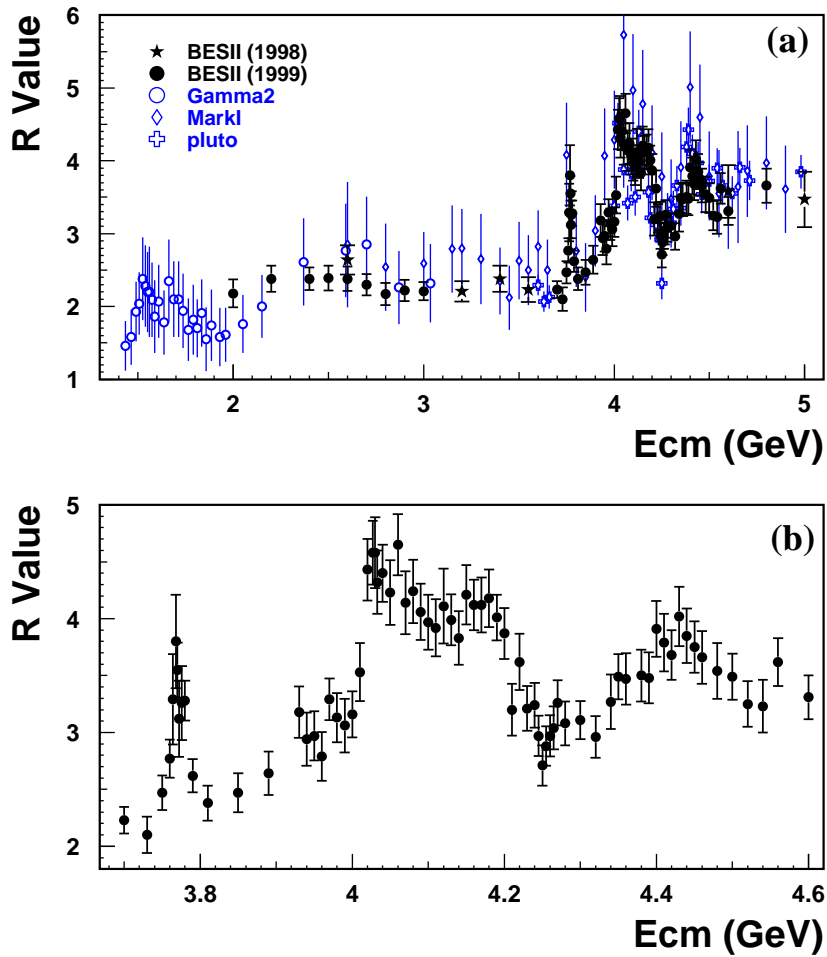}
\caption{Behavior of $R_{e^+ e^-}$ as a function of center-of-mass energy
(bottom plot shows magnified range) \cite{Bai:2001ct} showing sharp drop
at $\sqrt{s}=4260$ MeV just below S-wave charm-anticharm threshold (4285 MeV).
\label{fig:R}}
\end{figure}

\begin{figure}
\includegraphics[width=0.98\textwidth]{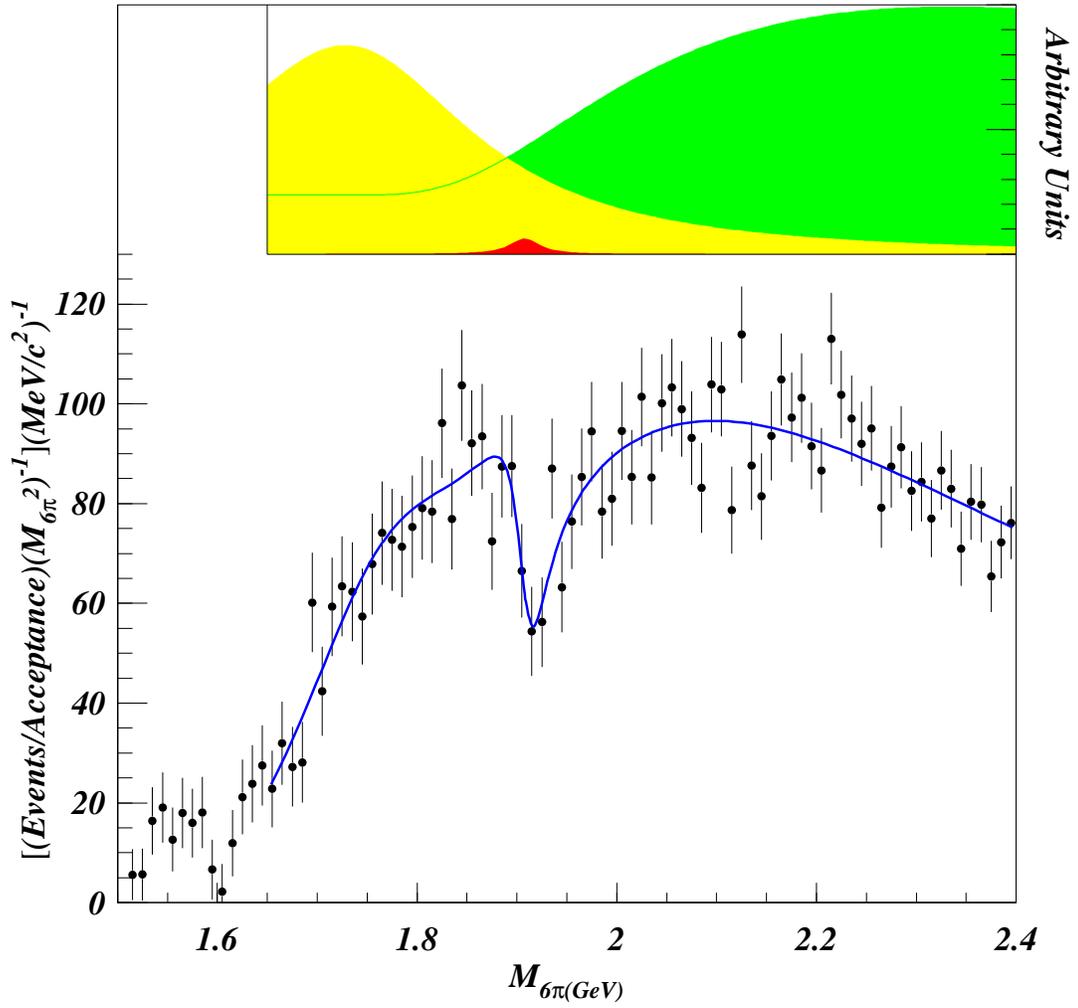}
\caption{Mass spectrum $M(3\pi^+ 3\pi^-)$ in photoproduction of six charged
pions \cite{Frabetti:2001ah,Frabetti:2003pw}.
\label{fig:6pic}}
\end{figure}

If an elastic phase shift goes though $180^\circ$, the scattering amplitude
vanishes: this is the {\it Ramsauer--Townsend effect} \cite{rte}.  It leads to
atomic or nuclear transparency at specific energies and can be utilized for
making monochromatic neutron beams \cite{Barbeau:2007qh}.  Sharp dips in mass
spectra, often correlated with S-wave thresholds, occur in many instances of
particle physics.  One example is the S-wave $\pi \pi$ spectrum near $K \bar K$
threshold.  The value of $R \equiv \sigma(e^+ e^- \to {\rm hadrons})/
\sigma(e^+ e^- \to \mu^+ \mu^-)$ drops sharply around $\sqrt{s} = 4.26$ GeV
\cite{Bai:2001ct} (see Fig.\ \ref{fig:R}), which happens to be just below the
threshold for production of ($D \bar D_1~+ $ charge conjugate), where $D$
and $D_1$ are charmed mesons with $J^{PC} = 0^-$ and $1^+$, respectively.
Diffractive photoproduction of $3 \pi^+ 3 \pi^-$  exhibits a dip near $p \bar
p$ threshold \cite{Frabetti:2001ah,Frabetti:2003pw} (Fig.\ \ref{fig:6pic}).

\section{Exotic baryonium? \label{sec:bary}}

\begin{figure}
\includegraphics[width=0.9\textwidth]{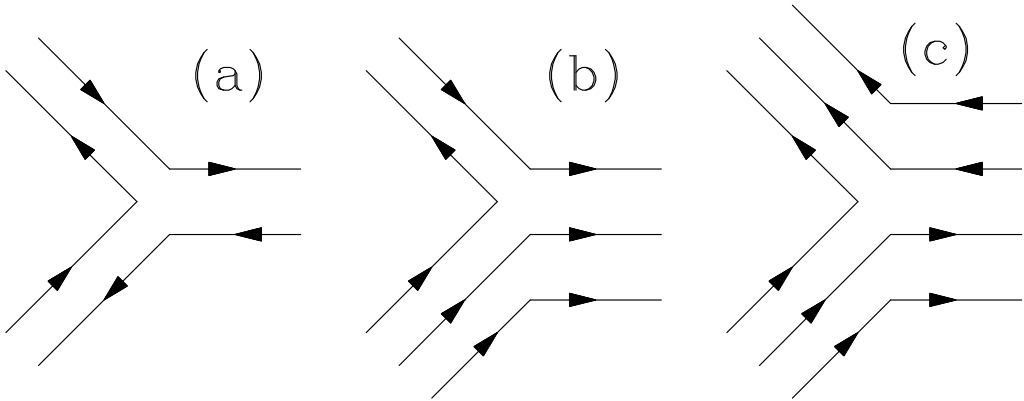}
\caption{Model for formation of resonances.  (a) Ordinary ($q \bar q$) meson
from two ordinary mesons; (b) Ordinary baryon ($qqq$) from
ordinary meson and ordinary baryon; (c) Exotic meson ($q q \bar q \bar q$)
from ordinary baryon and ordinary antibaryon.
\label{fig:form}}
\end{figure}

It has been suggested \cite{Rosner:1968si} that baryon-antibaryon states can
form exotic ($q q \bar q \bar q$) mesons,
as illustrated in Fig.\ \ref{fig:form}.  Indeed, if an ordinary meson contains
a quark $q_i$ and another meson contains an antiquark $\bar q_i$ of the same
flavor, they will form a resonance [Fig.\ 3(a)] when the center-of-mass (c.m.)
3-momentum typically does not exceed 350 MeV/$c$ \cite{Rosner:1973fq}.  The
corresponding c.m.\ 3-momentum for formation of a meson-baryon resonance
[Fig.\ 3(b)] is 250 MeV/$c$, and was estimated in Ref.\ \cite{Rosner:1973fq}
to be 200 MeV/$c$ for baryon-antibaryon resonance formation.

\begin{figure}
\includegraphics[width=0.3\textwidth]{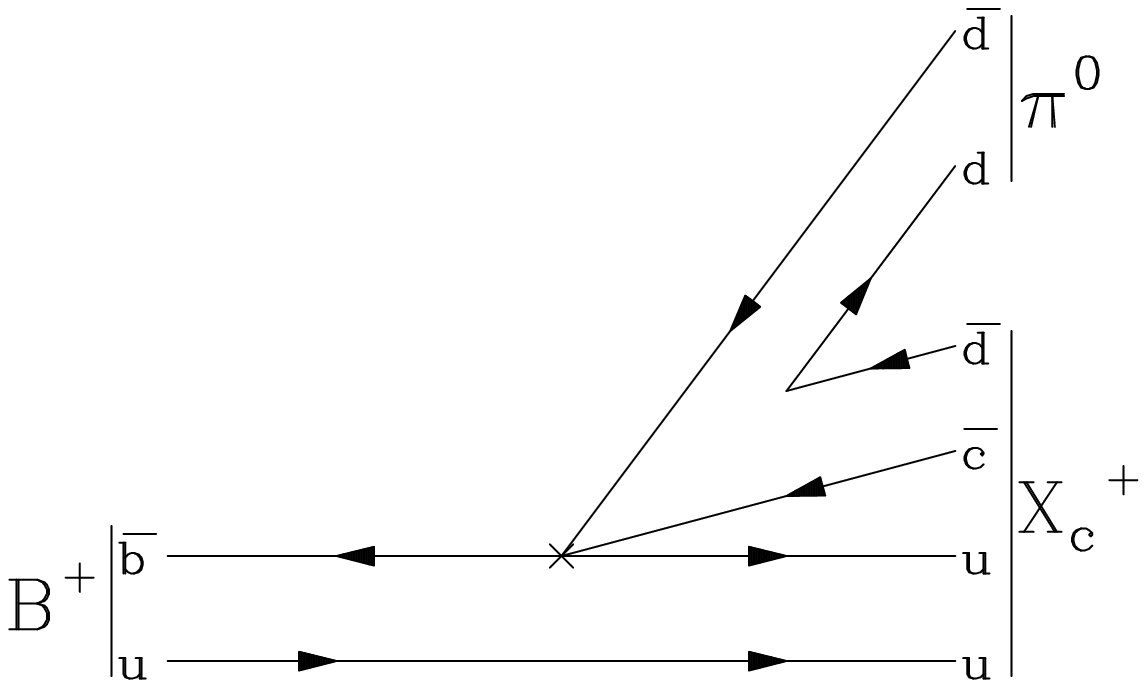}
\includegraphics[width=0.3\textwidth]{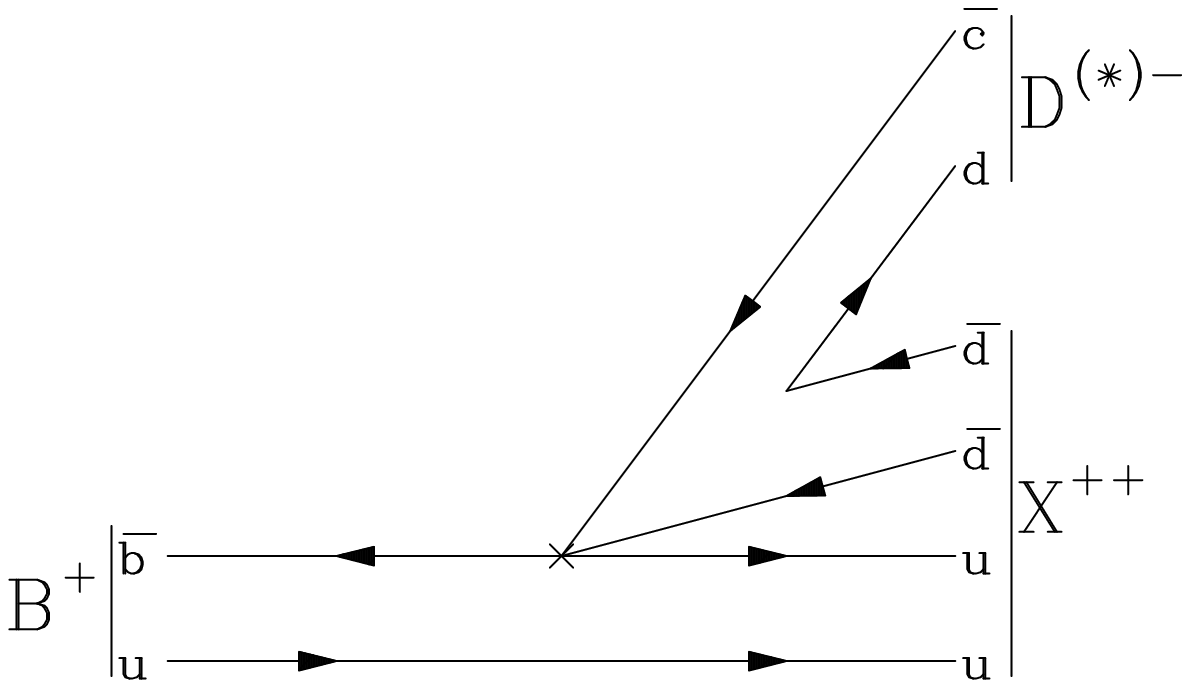}
\includegraphics[width=0.3\textwidth]{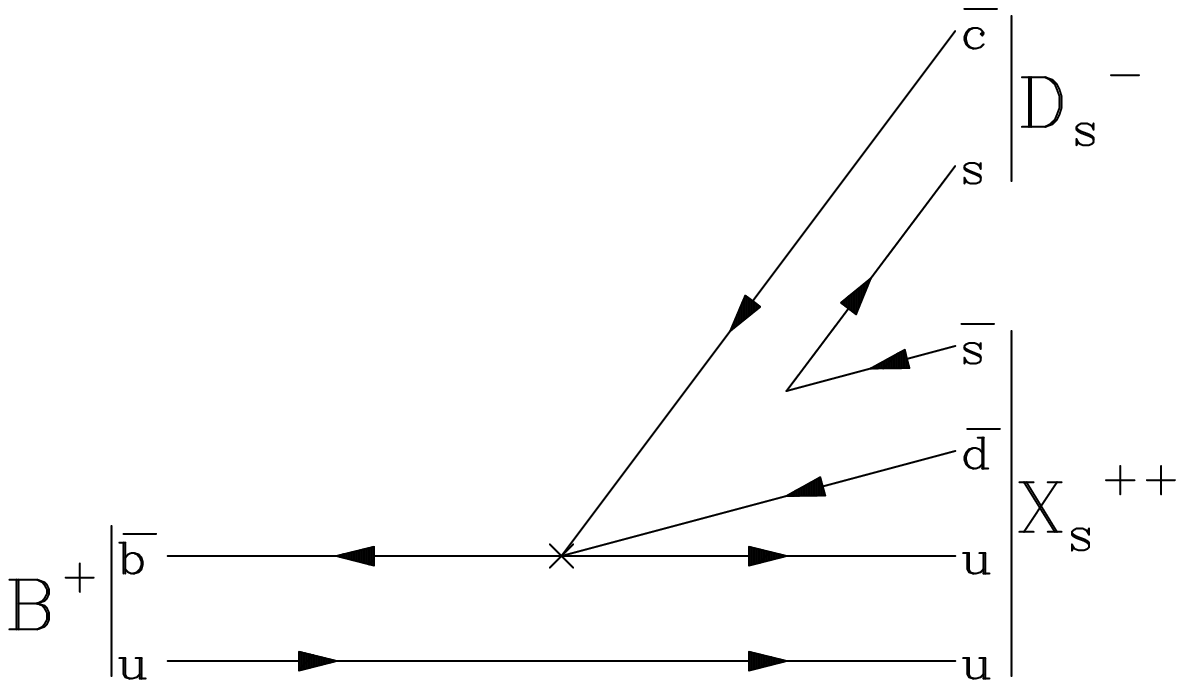}
\caption{Graphs depicting $B$ decays yielding flavor-exotic $q q \bar q \bar q$
mesons.  (a) $X_c^+ = u u \bar c \bar d$; (b) $X^{++} = uu \bar d \bar d$;
(c) $X^{++} = uu \bar d \bar s$
\label{fig:xpp}}
\end{figure}

A flavor state which cannot be formed of $q_1 \bar q_2$, such as $q_1 q_1
\bar q_2 \bar q_2$ is truly exotic. $B$ meson decays offer numerous exotic
final states, for example in $b \bar d \to c \bar u d \bar d$.  Examples of 
suggestions for seeing exotics at $B$ factories \cite{Rosner:2003ia,%
Terasaki:2011jt} are shown in Fig.\ \ref{fig:xpp}.

A bet was made with Peter Freund in 1972 that exotic baryonium would not be
found in two years (he bet it would).  He bought dinner in 1974; we are still
waiting for the discovery.  The decays of $B$ mesons can also yield
pentaquarks ($qqqq \bar q$ candidates \cite{Rosner:2003ia}); none have been
seen so far.

\section{Large $h_b$ production rate \label{sec:hb}}

Belle \cite{Adachi:2011ji,Bondar} has reported a large cross section for
$e^+ e^- \to \pi^+ \pi^- h_b(1P)$ or $\pi^+ \pi^- h'_b(2P)$ at the
center-of-mass energy of $\Upsilon(5S)$.  This is reminiscent of CLEO's
observation of a large cross section for $e^+ e^- \to \pi^+ \pi^- h_c$
at $\sqrt{s} = 4170$ MeV \cite{Pedlar:2011uqa,Rosner:2011}.  Earlier, BaBar
\cite{Aubert:2006bm,Aubert:2008bv} and Belle \cite{Sokolov:2006sd}
reported $\pi^+ \pi^-$ and $\eta$ transitions to lower $\Upsilon$ states from
$\Upsilon(4S)$ states; Belle \cite{Abe:2007tk} saw $\Gamma[\Upsilon(5S) \to
\pi^+ \pi^- \Upsilon(1S)] = (0.59 \pm 0.04 \pm 0.09)$ MeV, $\Gamma[\Upsilon(5S)
\to \pi^+ \pi^- \Upsilon(2S)] = (0.85 \pm 0.07 \pm 0.16)$ MeV, more than $10^2$
times the $nS$ rate for $n\le4$.

Some time ago Lipkin and Tuan \cite{Lipkin:1988tg} and Moxhay
\cite{Moxhay:1988ri} pointed out that rescattering from $B^{(*)} \bar B^{(*)}$
would be important in quarkonium production from states above flavor
threshold.  More recent calculations \cite{Meng:2007tk,Simonov:2008ci} support
this point of view, borne out by the prominence of peaks in $\pi h_b$ mass
spectra at $B \bar B^*$ and $B^* \bar B^*$ threshold
reported at this Conference \cite{Bondar}.  The masses of the $h_b(1P)$
and $h_b(2P)$ \cite{Adachi:2011ji,Bondar}, as well as the $h_c(1P)$ discovered
earlier, are very close to the spin-weighted averages of the corresponding
$^3P$ states, indicating small hyperfine splitting in $P$-wave mesons as
expected in the naive quark model.  Loop corrections from coupled channels
largely cancel and are found to be insignificant \cite{Burns:2011fu}.

\begin{figure}
\hskip 0.8in \includegraphics[width=0.7\textwidth]{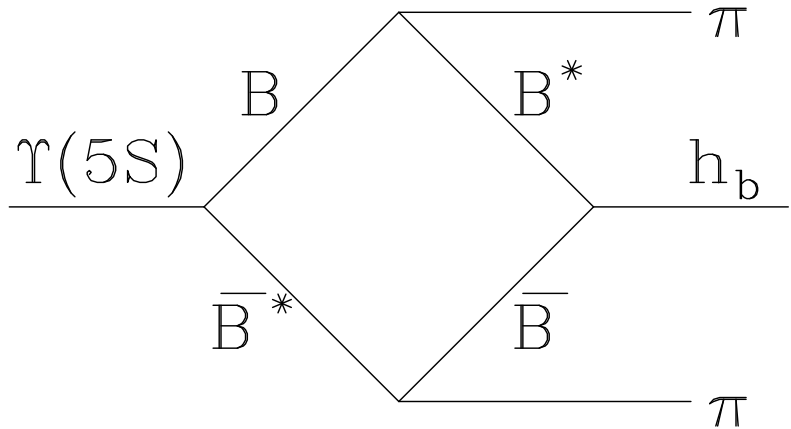}
\caption{One of several graphs in which rescattering from flavored
meson-antimeson pairs contributes to the process $\Upsilon(5S) \to \pi^+ \pi^-
h_b(1P,2P)$.
\label{fig:resc}}
\end{figure}

One of many graphs contributing to the rescattering process $\Upsilon(5S) \to B
\overline{B}^*,B^* \overline{B}^*,$ $\ldots \to \pi^+ \pi^- h_b$
\cite{Bugg:2011ub} is illustrated
in Fig.\ \ref{fig:resc}.  The energy must be above $B \overline{B}^*$
threshold in order to produce some $J^P(b \bar b)$ values.  A recent
description by Bondar {\it et al.} \cite{Bondar:2011ev} addresses selection
rules whereby certain bottomonium states are favorably produced in
rescattering.  Rescattering through flavored pairs flips the $b \bar b$ spin
from triplet to singlet in $\Upsilon(5S) \to \pi^+ \pi^- h_b(1P,2P)$, whereas
such a spin flip would be suppressed in perturbative QCD by an inverse power
of the bottom quark mass.

\section{Radiative transitions involving $\chi_b(1P)$ states \label{sec:radt}}

\begin{figure}
\hskip 0.2in
\includegraphics[width=0.92\textwidth]{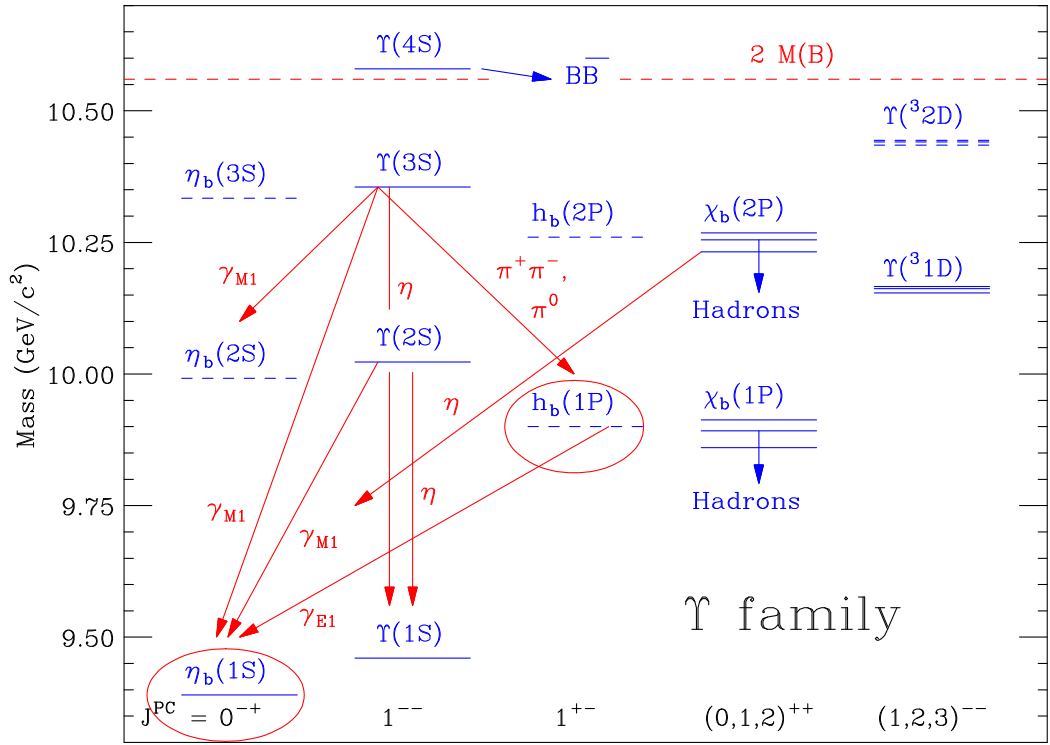}
\caption{Bottomonium states and transitions.  Not shown:  electric dipole
(E1) transitions between S and P states and between P and 1D states.
\label{fig:ups}}
\end{figure}

The lowest-lying states of the bottomonium spectrum are illustrated in Fig.\
\ref{fig:ups}.  We have heard about Belle's conclusive observation of the
$h_b(1P)$ and $h_b(2P)$ states \cite{Adachi:2011ji,Bondar}.  CLEO also
searched for the $h_b(1P)$, via the transitions $\Upsilon(3S) \to (\pi^+ \pi^-
h_b,~\pi^0 h_b,$ \\ $h_b \to \gamma \eta_b)$.  A significant background to the
$h_b$ search in the $\Upsilon(3S) \to \pi^0 h_b$ decay \cite{Ge:2011kq}
turned out to be the pairing
of a photon from $\Upsilon(3S) \to \gamma \chi_b(1P) \to \gamma \gamma
\Upsilon(1S)$, which required a more detailed study of these suppressed E1
transitions.  This section describes that investigation \cite{Kornicer:2010cb}.

Photons in the transitions $\Upsilon(3S) \to \gamma \chi_b(1P)$ and $\chi_b
\to \gamma \Upsilon(1S)$ are in the 400--500 MeV range and can be a problematic
background to the search for $\Upsilon(3S) \to \pi^0 h_b$.  The rates for
$\Upsilon(3S) \to \gamma \chi_b(1P)$, while small, are poorly known.  The
electric dipole matrix element between 3S and 1P states vanishes for a harmonic
oscillator potential and is highly suppressed for realistic quarkonium
potentials \cite{Grant:1995hf}.  Their values for various $\chi_{bJ}(1P)$
states thus test specific models of relativistic corrections, whose predictions
span a wide range.  Table \ref{tab:prev} summarizes previously known branching
fractions involving the $\chi_{bJ}(1P)$ states.  

\begin{table}
\caption{Previously known branching fractions involving $\chi_{bJ}(1P)$
bottomonium states \cite{Kornicer:2010cb}.  Where not shown otherwise, values
are taken from Ref.\ \cite{Nakamura:2010zzi}.
\label{tab:prev}}
\begin{center}
\begin{tabular}{|c|c|c|c|} \hline \hline
Transition ~~~ & $E_\gamma$ (MeV) & $\bf$ (\%) & Comments \\
\hline
$\Upsilon(3S) \to \gamma \chi_{b0}(1P)$~~~& 483.9 & $0.30 \pm 0.11$ & CLEO,
PR D {\bf 78}, 091103 \\ \cline{4-4}
$\Upsilon(3S) \to \gamma \chi_{b1}(1P)$    & 452.1 & $< 0.17$ & First reported
here \\ \cline{4-4}
$\Upsilon(3S) \to \gamma \chi_{b2}(1P)$  & 433.5 & $<1.9$ & First reported here
\\ \cline{4-4}
\hline
$\Upsilon(2S) \to \gamma \chi_{b0}(1P)$  & 162.5 & $3.8 \pm 0.4$ & Dominated by
CLEO: \\
$\Upsilon(2S) \to \gamma \chi_{b1}(1P)$  & 129.6 & $6.9 \pm 0.4$ & M. Artuso
{\it et al.}, \\
$\Upsilon(2S) \to \gamma \chi_{b2}(1P)$ & 110.4 & $7.15 \pm 0.35$ & PRL
{\bf 94}, 032001 (2005) \\
\hline
$\chi_{b0}(1P) \to \gamma \Upsilon(1S)$  & 391.1 & $< 6$ & Main $\chi_{b0}$
decay hadronic \\ \cline{4-4}
$\chi_{b1}(1P) \to \gamma \Upsilon(1S)$    & 423.0 & $35 \pm 8$ & Latest
measurement \\
$\chi_{b2}(1P) \to \gamma \Upsilon(1S)$  & 441.6 & $22 \pm 4$ & in 1986! \\
\hline \hline
\end{tabular}
\end{center}
\end{table}

\subsection{Unfolding 420--450 MeV photons \label{subsec:unf}} 

The overlap of photon energies illustrated in Table \ref{tab:prev} means it
is easiest to quote the summed combination of branching fractions
\beq
\brf_{\rm{sum}}= \sum_{J=1,2} \brf[\Upsilon(3S) \to \gamma \chi_{bJ}(1P)]\times
\brf[\chi_{bJ}(1P) \to \gamma \Upsilon(1S)]
\eeq
$= (1.2^{+0.4}_{-0.3} \pm 0.09) \times 10^{-3}$ \cite{Heintz:1992cv}
or $(2.14 \pm 0.22 \pm 0.21) \times 10^{-3}$ \cite{Skwarnicki:2002bp}

\begin{figure}
\includegraphics[width=0.47\textwidth]{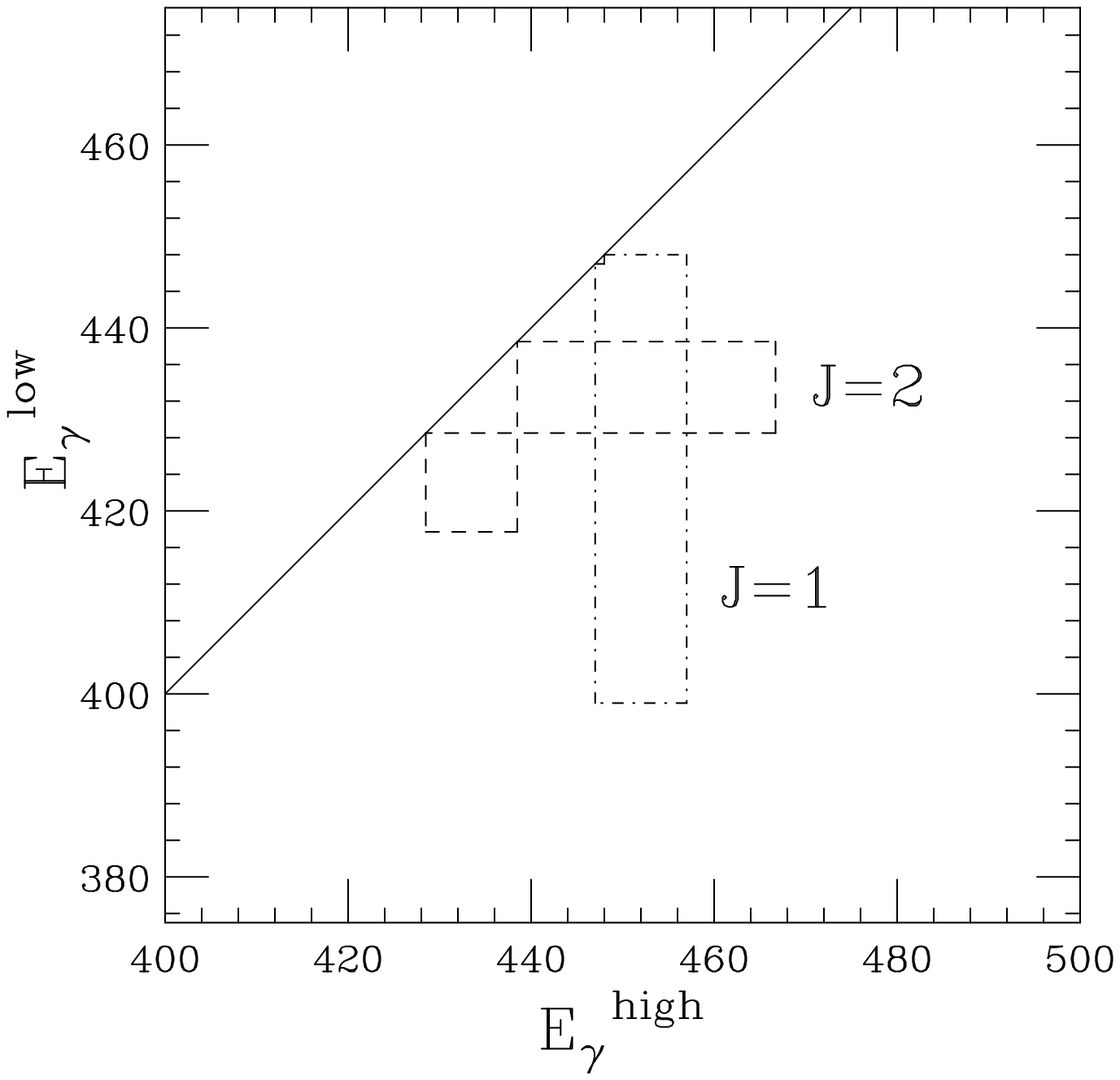}
\includegraphics[width=0.47\textwidth]{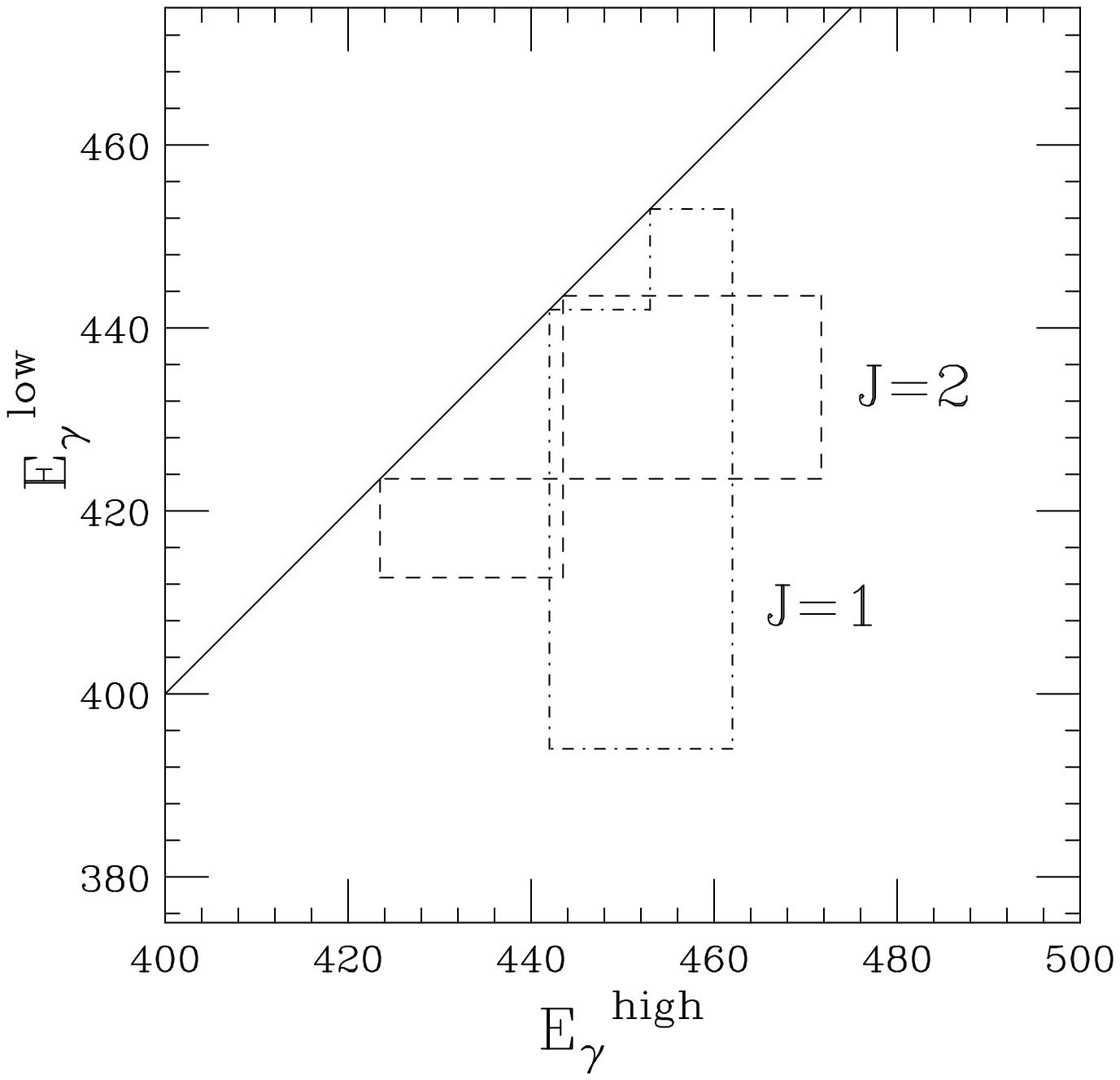}
\caption{Energy of lower-energy vs.\ higher-energy photon in the transitions
$\Upsilon(3S) \to \gamma \chi_{bJ} \to \gamma \gamma \Upsilon(1S)$ for
$J=1$ and $J=2$.  Photon energy spread is taken to be $\pm$5 MeV (left) or
$\pm$10 MeV (right).
\label{fig:ehilo}}
\end{figure}

\begin{figure}
\includegraphics[width=0.43\textwidth]{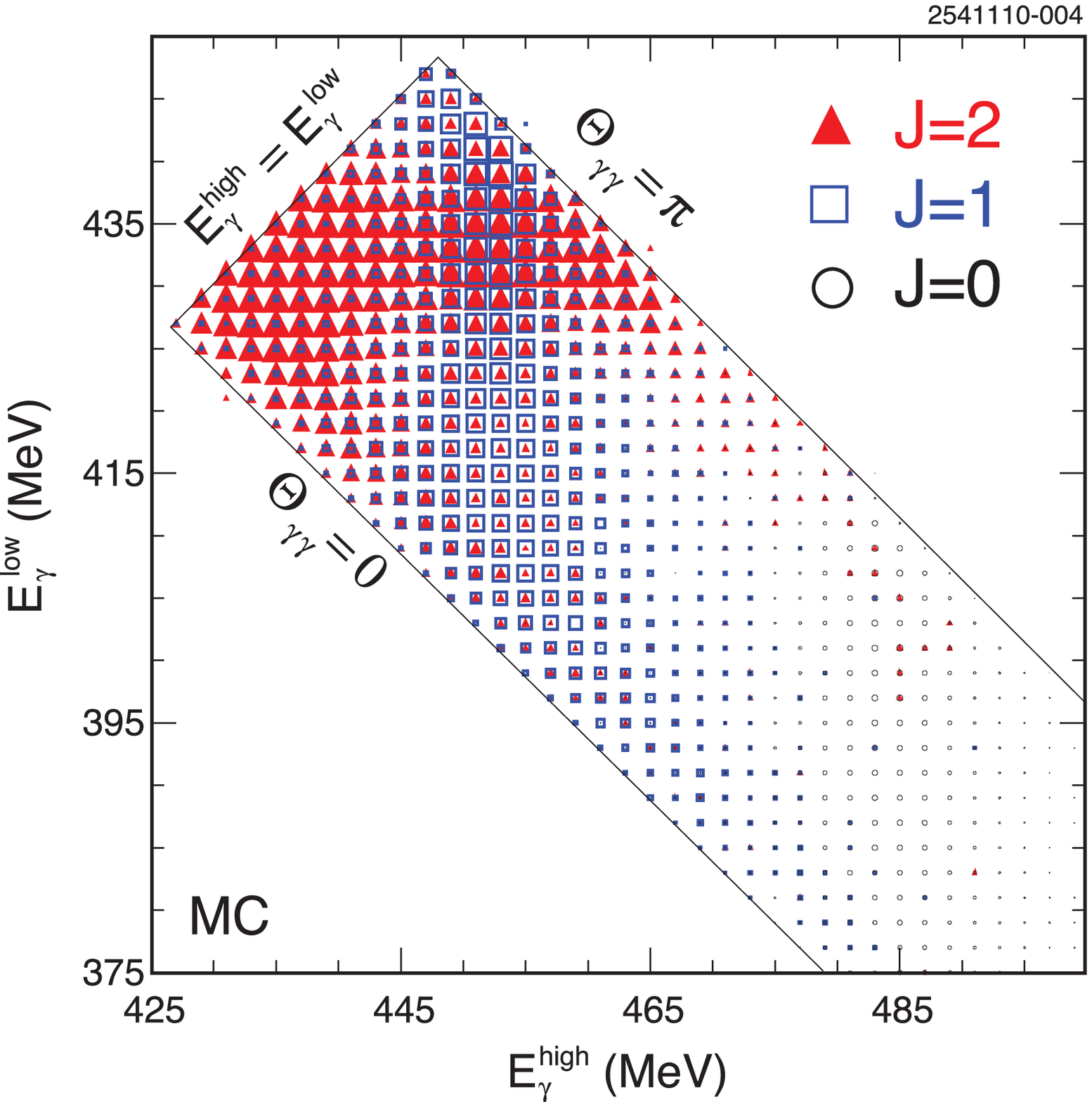}
\includegraphics[width=0.45\textwidth]{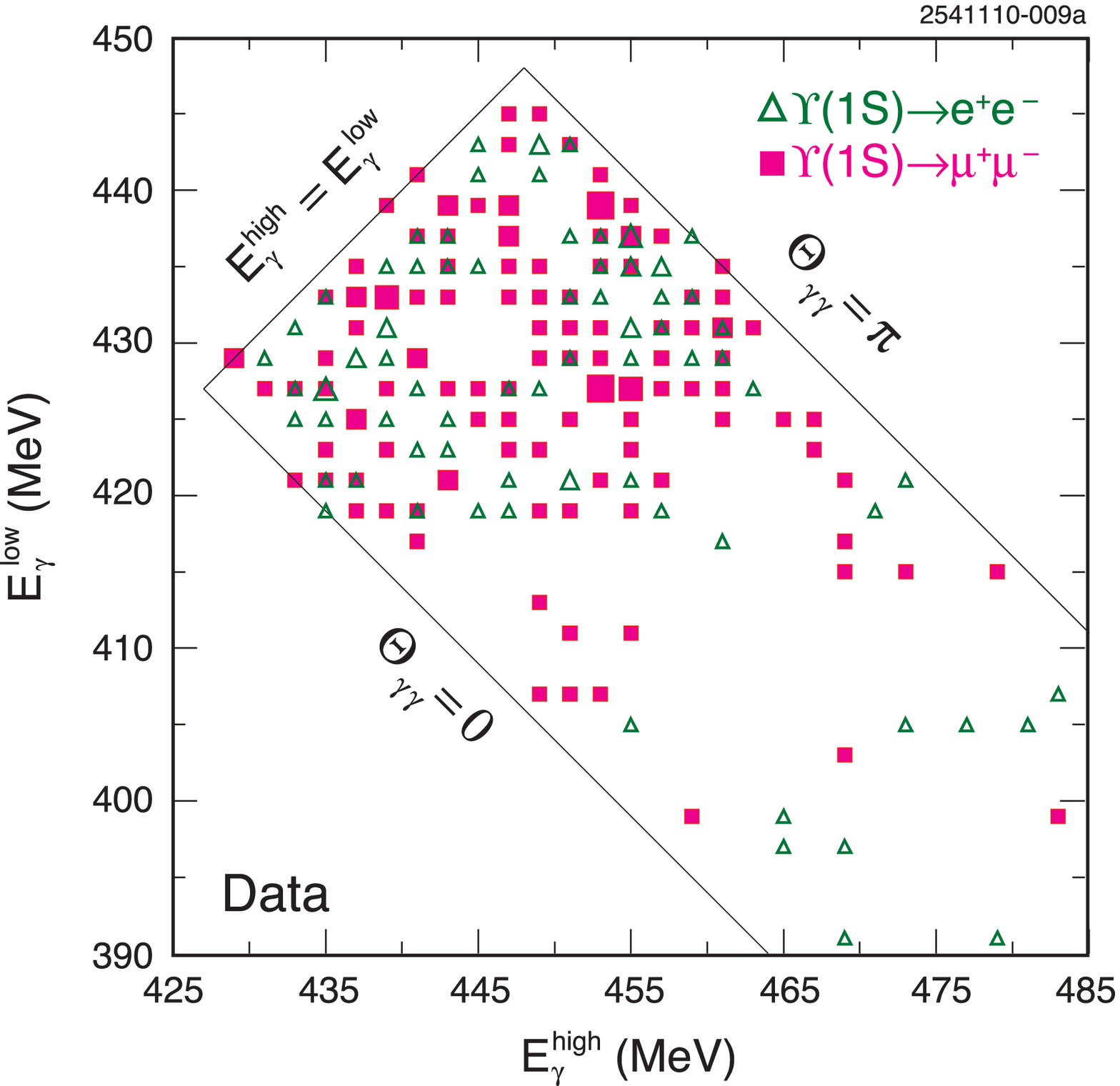}
\caption{Monte Carlo simulation (left) and data (right) for distributions of
$E_\gamma^{\rm high}$ vs.\ $E_\gamma^{\rm low}$ in transitions $\Upsilon(3S)
\to \gamma \chi_{bJ} \to \gamma \gamma \Upsilon(1S)$.  In the left-hand panel,
$\Upsilon(1S) \to \mu^+ \mu^-$; the distribution for $\Upsilon(1S) \to e^+ e^-$
is similar.  In the right-hand panel, triangles correspond to
$\Upsilon(1S) \to e^+ e^-$, while boxes correspond to $\Upsilon(1S)
\to \mu^+ \mu^-$.
\label{fig:mcd}}
\end{figure}

To unfold the $J=1$ and $J=2$ contributions one may use Doppler broadening,
as illustrated in Fig.\ \ref{fig:ehilo}.  Here we have plotted the expected
energies of the lower- vs.\ higher-energy photon in the transitions
$\Upsilon(3S) \to \gamma \chi_{bJ} \to \gamma \gamma \Upsilon(1S)$ for
$J=1$ and $J=2$ under two different assumptions about the photon energy spread.
One sees that even with a $\pm$10 MeV energy spread, the transitions involving
$J=1$ and $J=2$ states populate different regions of the $E_\gamma^{\rm high}$%
--$E_\gamma^{\rm low}$ plane.  This is borne out by a Monte Carlo simulation
(Fig.\ \ref{fig:mcd}, left panel).

A two-dimensional fit to the data (right-hand panel of
Fig.\ \ref{fig:mcd}) provides the best sensitivity to the separate $J=1$
and $J=2$ components (the $J=0$ contribution is negligible because of its
small branching fraction to $\gamma \Upsilon(1S)$).  We define
\beq
\brf1 \equiv \brf[\Upsilon(3S) \to \gamma \chi_{bJ}(1P)],~
\brf2 \equiv \brf[\chi_{bJ}(1P) \to \gamma \Upsilon(1S)],~
\brf3 \equiv \brf[\Upsilon(1S) \to \ell^+ \ell^-].
\eeq
We take $\brf2($J=1$) = (33.0 \pm 0.5)\%$ and $\brf2($J=2$) = (18.5 \pm 0.5)\%$
from a new fit to $\Upsilon(2S)$ data \cite{Kornicer:2010cb}, and $\brf3 =
(2.48 \pm 0.05)\%$ \cite{Nakamura:2010zzi} assuming muon-electron universality.
For the sum of the $J=1$ and $J=2$ contributions, we find $\sum \brf1 \times
\brf2 = (2.00 \pm 0.15 \pm 0.22 \pm 0.04) \times 10^{-3}$, agreeing well with
the 2002 CLEO value \cite{Skwarnicki:2002bp}.  Determinations for individual
values of $J$ are summarized in Table \ref{tab:each}.  Portions of Table
\ref{tab:prev} now are changed to those summarized in Table \ref{tab:now}.
Also shown are new values from BaBar using converted photons \cite{Lees:2011mx}.

\begin{table}
\caption{Branching fractions $\brf1 \times \brf2$ and $\brf1$, where
$\brf1 \equiv \brf[\Upsilon(3S) \to \gamma \chi_{bJ}(1P)$ and $\brf2 \equiv
\brf[\chi_{bJ}(1P) \to \gamma \Upsilon(1S)]$, for individual values of $J$.
\label{tab:each}}
\begin{center}
\begin{tabular}{c c c} \hline \hline
 & $J=1$ & $J=2$ \\ \hline
$\brf1 \times \brf2~(10^{-4})$ & $5.38 \pm 1.20 \pm 0.94 \pm 0.11$
    & $14.35 \pm 1.62 \pm 1.66 \pm 0.29$ \\
$\brf1~(10^{-3})$ & $1.63 \pm 0.36 \pm 0.28 \pm 0.09$
    & $7.74 \pm0.88 \pm0.88 \pm0.38$ \\ \hline \hline
\end{tabular}
\end{center}
\end{table}

\begin{table}
\caption{Branching fractions of Table \ref{tab:prev} updated in the present
analysis \cite{Kornicer:2010cb}.
\label{tab:now}}
\begin{center}
\begin{tabular}{c c c c} \hline \hline
Transition ~~~ & \multicolumn{3}{c}{$\brf$ (\%)} \\
               & Previous & CLEO now & Babar \cite{Lees:2011mx} \\ \hline
$\Upsilon(3S) \to \gamma \chi_{b0}(1P)$ & $0.30 \pm 0.11$ & $0.30 \pm 0.11$
 & $0.27 \pm 0.04$ \\
$\Upsilon(3S) \to \gamma \chi_{b1}(1P)$ & $< 0.17$ & $0.163 \pm 0.046$
 & $0.05^{+0.04}_{-0.03}~(< 1.1)$ \\
$\Upsilon(3S) \to \gamma \chi_{b2}(1P)$ & $<1.9$ & $0.774 \pm 0.130$
 & $1.06 \pm 0.07$ \\ \hline
$\chi_{b0}(1P) \to \gamma \Upsilon(1S)$ & $< 6$ & $1.73 \pm 0.35$
 & $2.3^{+1.8}_{-1.7}~(<4.6)$ \\
$\chi_{b1}(1P) \to \gamma \Upsilon(1S)$ & $35 \pm 8$ & $33.0 \pm 2.6$
 & $36.2 \pm 2.8$ \\
$\chi_{b2}(1P) \to \gamma \Upsilon(1S)$ & $22 \pm 4$ & $18.5 \pm 1.4$
 & $20.2^{+1.6}_{-1.9}$ \\
\hline \hline
\end{tabular}
\end{center}
\end{table}

\subsection{Experiment vs.\ theory for $\Upsilon(3S) \to \gamma \chi_{bJ}$
\label{subsec:comp}}

\begin{figure}
\includegraphics[width=0.98\textwidth]{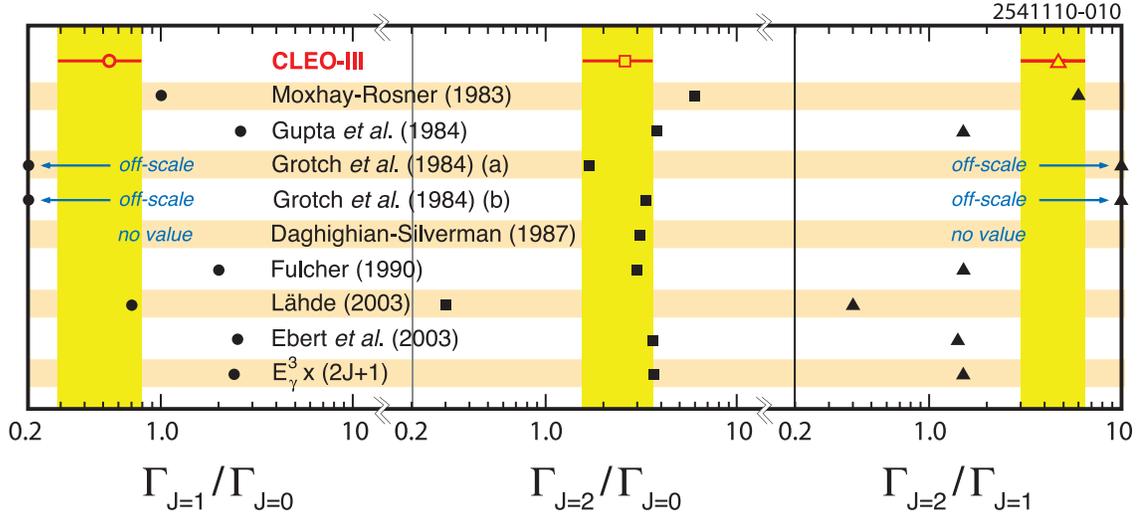}
\caption{Comparison of measured and predicted ratios of rates
for $\Upsilon(3S) \to \gamma \chi_{bJ}$.
\label{fig:comp}}
\end{figure}

In Table \ref{tab:comp} we compare measured partial widths for the transitions
$\Upsilon(3S) \to \gamma \chi_{bJ}$, including a previous measurement of
$\Gamma_{J=0}$ in an inclusive CLEO experiment \cite{upsart}, with a number of
theoretical predictions \cite{thmodels}.  Fig.\ \ref{fig:comp} compares
measured and predicted ratios of these rates.  More significant than the
agreement with any one model is the spread in predictions (note the log scale
in Fig.\ \ref{fig:comp}!), and the observation
that the $\Upsilon(3S) \to \gamma \chi_{bJ}(1P)$ rates differ from the pattern
$\sim E_\gamma^3 (2J + 1)$ expected in a nonrelativistic approach.

\begin{table}
\caption{Comparison of measured and predicted values of $\Gamma[\Upsilon(3S)
\to \gamma \chi_{bJ}(1P)]$.
\label{tab:comp}}
\begin{center}
\begin{tabular}{c c c c}
 &$\Gamma_{J=0}$ (eV) & $\Gamma_{J=1}$ (eV) & $\Gamma_{J=2}$ (eV) \\ \hline
This analysis & -- & $33\pm10$ & $157\pm30$ \\
Inclusive CLEO expt. & $61\pm23$ & -- & -- \\ \hline
Moxhay--Rosner (1983) & $25$ & $25$ & $150$ \\
Gupta {\it et al.} (1984) & $1.2$ & $3.1$ & $4.6$ \\
Grotch {\it et al.} (1984) (a) & $114$ & $3.4$ & $194$ \\
Grotch {\it et al.} (1984) (b) & $130$ & $0.3$ & $430$ \\
Daghighian--Silverman (1987) & $42$ & -- & $130$ \\
Fulcher (1990) & $10$ & $20$ & $30$ \\
L\"ahde (2003) & $150$ & $110$ & $40$ \\
Ebert {\it et al.} (2003) & $27$ & $67$ & $97$ \\ \hline
\end{tabular}
\end{center}
\leftline{(a) Scalar confining potential. (b) Vector confining potential.}
\end{table}

The deviations from the nonrelativistic pattrern of partial widths
$\sim E_\gamma^3(2J+1)$ tests models of relativistic corrections.  It is
probably worth revisiting some of the old calculations within newer frameworks,
such as NRQCD.  We may also compare new results for the branching fractions
${\cal B}[\chi_{bJ}(1P) \to \gamma \Upsilon(1S)]$ with theoretical predictions
\cite{thmodels,KR}; see Table \ref{tab:comp1} \cite{Kornicer:2010cb}.

\begin{table}
\caption{Comparison of results for ${\cal B}[\chi_{bJ}(1P) \to \gamma
\Upsilon(1S)]$ with theoretical predictions (in \%) \cite{thmodels,KR}.
\label{tab:comp1}}
\begin{center}
\begin{tabular}{c c c c} \hline
Reference & $J=0$ & $J=1$ & $J=2$ \\ \hline
CLEO-III & $1.73\pm0.35$ & $33.0\pm2.6$ & $18.3\pm1.4$ \\
Moxhay--Rosner (1983)  & 3.8 & 50.6 & 22.3 \\
Gupta {\it et al.} (1984) & 4.1 & 56.8 & 26.7 \\
Grotch {\it et al.} (1984) (a) & 3.1 & 41.9 & 19.4 \\
Grotch {\it et al.} (1984) (b) & 3.3 & 43.9 & 20.3 \\
Daghighian--Silverman (1987) & 2.3 & 31.6 & 16.6 \\
Kwong--Rosner (1988) & 3.2 & 46.1 & 22.2 \\
Fulcher (1990) & 3.1 & 39.9 & 18.6 \\
L\"ahde (2003) & 3.3 & 45.7 & 21.1 \\
Ebert {\it et al.} (2003) & 3.7 & 51.5 & 23.6 \\ \hline
\end{tabular}
\end{center}
\leftline{(a) Scalar confining potential. (b) Vector confining potential.}
\end{table}

Most of the predicted branching fractions for the electric dipole transitions
in Table \ref{tab:comp1} are systematically larger than the experimental
values, indicating that the hadronic widths $\Gamma_h$ were underestimated.
An increase in the assumed value of $\alpha_S(m_b^2)$ leads to better agreement
with experiment.  For example, the values in Ref.\ \cite{KR} were calculated
for $\alpha_S(m_b^2) = 0.18$.  Using dependence on $\alpha_S$ of hadronic
widths for the $\chi_{bJ}$ states \cite{KMRR}, an increase of $\alpha_S(m_b^2)$
to $0.214 \pm 0.006$ leads to a satisfactory description of the branching
fractions, and is consistent with a recent compilation \cite{Bethke:2009jm}.

\section{Conclusions \label{sec:concl}}

Heavy quarkonium theory now must confront light-quark degrees of freedom.
Although we have been living with this since the dawn of hadron spectroscopy,
new experimental results reinforce the viewpoint that mesonic degrees of
freedom are important.  Scalar mesons' properties are governed by the $\pi
\pi$, $K \pi$, and $K \bar K$ channels to which they couple.  Effects of S-wave
thresholds are ubiquitous.

We are still waiting for definitive evidence for tetraquark exotics.  The
recent discoveries by the Belle Collaboration of enhanced $h_b(1P)$ and
$h_b(2P)$ production in $\Upsilon(5S) \to \pi^+ \pi^- h_b(1P,2P)$
\cite{Adachi:2011ji,Bondar} and of prominent enhancements of the $\pi
h_b$ mass spectra at $B \overline{B}^*$ and $B^* \overline{B}^*$ thresholds
\cite{Bondar} serve as a challenge to our understanding of hadron interactions,
but indicate a key role for rescattering from flavored meson-antimeson
intermediate states \cite{Bugg:2011ub,Bondar:2011ev}.

Finally, progress in the study of bottomonium electromagnetic transitions has
provided new data with which to confront models of relativstic corrections
to na\"{\i}ve quarkonium pictures.  We look forward to such calculations
on a firmer footing.

\Acknowledgements
This work was supported in part by the United States Department of Energy
under Grant No.\ DE-FG02-90ER40560.

\end{document}